\documentclass{emulateapj}
\usepackage{graphicx}
\usepackage{hyperref}
\usepackage{color}
\usepackage{amsmath}

\newcommand{\ergs}{${\rm erg \ cm^{-2} \ s^{-1}}$ }

\newcommand{\todo}{\ifmmode {\Huge \bullet} \else {\Huge$\bullet$}\fi}

\newcommand{\vFWHM}{\ifmmode V_{\mbox{\tiny FWHM}} \else $V_{\mbox{\tiny FWHM}}$ \fi}

\newcommand{\kms}{\ifmmode {\rm km\,s}^{-1} \else km\,s$^{-1}$ \fi}
\newcommand{\cmii}{\hbox{cm$^{-2}$}}
\newcommand{\cmiii}{\hbox{cm$^{-3}$}}
\newcommand{\ergcms}{\ifmmode {\rm ergs\,cm}^{-2}\,{\rm s}^{-1} \else ergs\,cm$^{-2}$\,s$^{-1}$\fi}
\newcommand{\ergcmsA}{\ifmmode{\rm ergs}\, {\rm cm}^{-2}\,{\rm s}^{-1}\,{\rm\AA}^{-1} \else ergs\, cm$^{-2}$\, s$^{-1}$\, \AA$^{-1}$\fi}
\newcommand{\ergcmsHz}{\ifmmode{\rm ergs\,cm}^{-2}\,{\rm s}^{-1}\,{\rm Hz}^{-1} \else ergs\,cm$^{-2}$\,s$^{-1}$\,Hz$^{-1}$\fi}
\newcommand{\phcms}{\ifmmode {\rm ph\,cm}^{-2}\,{\rm s}^{-1} \else ,ph\,cm$^{-2}$\,s$^{-1}$\fi}
\newcommand{\phcmsA}{\ifmmode {\rm ph\,cm}^{-2}\,{\rm s}^{-1}\,{\rm\AA}^{-1} \else ph\,cm$^{-2}$\,s$^{-1}$\,\AA$^{-1}$\fi}

\newcommand\Msun{\ifmmode M_{\odot} \else $M_{\odot}$\fi}
\newcommand\msun{\ifmmode M_{\odot} \else $M_{\odot}$\fi}
\newcommand\Lsun{\ifmmode L_{\odot} \else $L_{\odot}$\fi}

\newcommand\mpyr{\ifmmode \Msun\,{\rm yr}^{-1} \else $\Msun\,{\rm yr}^{-1}$ \fi}

\newcommand{\Luv}{\ifmmode L_{1450} \else $L_{1450}$\fi}
\newcommand{\Lop}{\ifmmode L_{5100} \else $L_{5100}$\fi}
\newcommand{\Lthree}{\ifmmode L_{3000} \else $L_{3000}$\fi}
\newcommand{\lledd}{\ifmmode L/L_{\rm Edd} \else $L/L_{\rm Edd}$\fi}
\newcommand{\ledd}{\ifmmode L_{\rm Edd} \else $L_{\rm Edd}$\fi}
\newcommand{\lamLlam}{\ifmmode \lambda L_{\lambda} \else $\lambda L_{\lambda}$\fi}
\newcommand{\lbol} {\ifmmode L_{\rm Bol} \else $L_{\rm Bol}$\fi}
\newcommand{\llbol}{\ifmmode \log\left(\lbol/\ergs\right) \else $\log\left(\lbol/\ergs\right)$\fi}

\newcommand{\fuv}{\ifmmode f_{\lambda}\left(1450\AA\right) \else $f_{\lambda}\left(1450 {\rm \AA}\right)$\fi}
\newcommand{\fthree}{\ifmmode f_{\lambda}\left(3000\AA\right) \else $f_{\lambda}\left(3000{\rm \AA}\right)$\fi}
\newcommand{\fH}{\ifmmode f_{\lambda}\left(1.65\micron\right) \else
$f_{\lambda}\left(1.65\micron\right)$\fi}

\newcommand{\mbh}{\ifmmode M_{\rm BH} \else $M_{\rm BH}$\fi}
\newcommand{\lmbh}{\ifmmode \log\left(\mbh/\Msun\right) \else $\log\left(\mbh/\Msun\right)$\fi}

\newcommand{\mseed}{\ifmmode M_{\rm seed} \else $M_{\rm seed}$\fi}
\newcommand{\mbul}{\ifmmode M_{\rm Bulge} \else $M_{\rm Bulge}$\fi}
\newcommand{\mstar}{\ifmmode M_{*} \else $M_{*}$\fi}
\newcommand{\mhost}{\ifmmode M_{\rm Host} \else $M_{\rm Host}$\fi}
\newcommand{\mm}{\ifmmode M_{*}/M_{\rm BH} \else $M_{*}/M_{\rm BH}$\fi}
\newcommand{\mmwp}{\ifmmode \left(M_{*}/M_{\rm BH}\right) \else $\left(M_{*}/M_{\rm BH}\right)$\fi}
\newcommand{\ml}{\ifmmode M_{*}/L_{*} \else $M_{*}/L_{*}$\fi}
\newcommand{\mlwp}{\ifmmode \left(M_{*}/L\right) \else $\left(M_{*}/L\right)$\fi}
\newcommand{\mlk}{\ifmmode \left(M_{*}/L_{K}\right) \else $\left(M_{*}/L_{K}\right)$\fi}
\newcommand{\sigs}{\ifmmode \sigma_{*} \else $\sigma_{*}$\fi}
\newcommand{\fbol} {\ifmmode f_{\rm bol} \else $f_{\rm bol}$\fi}
\newcommand{\fbolwv} {\ifmmode f_{\rm bol}\left(\lambda\right) \else $f_{\rm bol}\left(\lambda\right)$\fi}
\newcommand{\fbolopt} {\ifmmode f_{\rm bol}\left(5100\AA\right) \else $f_{\rm bol}\left(5100\AA\right)$\fi}
\newcommand{\fboluv} {\ifmmode f_{\rm bol}\left(3000\AA\right) \else $f_{\rm bol}\left(3000\AA\right)$\fi}

\newcommand{\zfpe}{\ifmmode z\simeq4.8 \else $z\simeq4.8$\fi}

\newcommand \Hbeta {\ifmmode {\rm H}\beta \else H$\beta$\fi}
\newcommand \hb    {\ifmmode {\rm H}\beta \else H$\beta$\fi}
\newcommand  \mgii  {\ifmmode {\rm Mg}{\textsc{ii}} \else Mg\,{\sc ii}\fi}
\newcommand  \MgII  {\ifmmode {\rm Mg}\,{\sc ii}\,\lambda2798 \else Mg\,{\sc ii}\,$\lambda2798$\fi}
\newcommand  \siiv  {\ifmmode {\rm Si}{\textsc{iv}} \else Si\,{\sc iv}\fi}
\newcommand  \SIIV  {\ifmmode {\rm Si}\,{\sc iv}\,\lambda1399 \else Si\,{\sc iv}\,$\lambda1399$\fi}
\newcommand  \civ  {\ifmmode {\rm C}\, {\sc iv}\ \else C\,{\sc iv}\fi}
\newcommand  \CIV  {\ifmmode {\rm C}\,{\sc iv}\,\lambda1549 \else C\,{\sc iv}\,$\lambda1549$\fi}
\newcommand  \NV  {\ifmmode {\rm N}\,{\sc v}\,\lambda1240 \else N\,{\sc v}\,$\lambda1240$\fi}
\newcommand  \nv  {\ifmmode {\rm N}\,{\sc v}\ \else N\,{\sc v}\fi}
\newcommand  \LyA  {\ifmmode {\rm Lyman}\,{\sc $\alpha$}\,\lambda1216 \else Lyman\,{\sc $\alpha$}\,$\lambda1216$\fi}
\newcommand  \lya {\ifmmode {\rm Lyman}\,{\sc $\alpha$}\ \else Lyman\,{\sc $\alpha$}\fi}
\newcommand  \feii     {\ifmmode {\rm Fe}\,{\sc II}\,\lambda1785.4 \else Fe\,{\sc II}\,$\lambda1785.4$\fi}

\newcommand  \aliii  {\ifmmode {\rm Al}{\textsc{iii}} \else Al\,{\sc iii}\fi}

\newcommand  \oi    {\ifmmode \left[{\rm O}\,{\textsc i}\right] \else [O\,{\sc i}]\fi}
\newcommand  \OI    {\ifmmode \left[{\rm O}\,{\textsc i}\right]\,\lambda6300 \else [O\,{\sc i}]$\,\lambda6300$ \fi}

\newcommand  \oii   {\ifmmode \left[{\rm O}\,{\textsc ii}\right] \else [O\,{\sc ii}]\fi}
\newcommand  \OII   {\ifmmode \left[{\rm O}\,{\textsc ii}\right]\,\lambda3727 \else [O\,{\sc ii}]\,$\lambda3727$ \fi}
\newcommand  \oiii  {\ifmmode \left[{\rm O}\,{\textsc iii}\right] \else [O\,{\sc iii}]\fi}
\newcommand  \OIII  {\ifmmode \left[{\rm O}\,{\textsc iii}\right]\,\lambda5007 \else [O\,{\sc iii}]\,$\lambda5007$\fi}

\newcommand{\lmg}{\ifmmode L\left(\mgii\right) \else $L\left(\mgii\right)$\fi}
\newcommand{\fwmg}{\ifmmode {\rm FWHM}\left(\mgii\right) \else FWHM(\mgii)\fi}
\newcommand{\fwciv}{\ifmmode {\rm FWHM}\left(\civ\right) \else FWHM(\civ)\fi}
\newcommand{\fwhm}{\ifmmode {\rm FWHM} \else FWHM\fi}

\begin{document}
\title{Leaked \lya~emission: an indicator of the size of quasar absorption outflows}
\author{Zhicheng He\altaffilmark{$\star$}, Guilin Liu\altaffilmark{$\dag$}, Tinggui Wang\altaffilmark{}, Chenwei Yang\altaffilmark{} and Zhenfeng Sheng\altaffilmark{}} 
\altaffiltext{}{
CAS Key Laboratory for Research in Galaxies and Cosmology, Department of 
Astronomy, University of Science and Technology of China, Hefei, Anhui 
230026; $^{\star}$zcho@mail.ustc.edu.cn; $^{\dag}$glliu@ustc.edu.cn}
\begin{abstract}
The galactocentric distance of quasar absorption outflows are conventionally determined using
absorption troughs from excited states, a method hindered by severely saturated or self-blended
absorption troughs. We propose a novel method to estimate the size of a broad absorption line (BAL)
region which partly obscures an emission line region by assuming virialized gas in the emission region
surrounding a supermassive black hole with known mass. When a spiky \LyA~line emission
is present at the flat bottom of the deep \NV~absorption trough, the size of BAL
region can be estimated. We have found 3 BAL quasars in the SDSS database showing such
\lya~lines. The scale of their BAL outflows are found to be 3-26 pc, moderately larger than
the theoretical scale (0.01-0.1pc) of trough forming regions for winds originating from accretion
discs, but significantly smaller than most outflow sizes derived using 
the absorption troughs of the excited states of ions. For these three outflows, the lower limits of ratio
of kinetic luminosity to Eddington luminosity are 0.02\%-0.07\%. 
These lower limits are substantially smaller than that is required to have significant 
feedback effect on their host galaxies.
\end{abstract}


\section{Introduction}
\label{sec:intro}

Quasar outflow, as an essential component of the quasar structure,
has been routinely invoked as a primary feedback mechanism to explain 
the growth of super-massive black holes (SMBHs), evolution of the host 
galaxies, enrichment of the intergalactic medium (IGM), cluster cooling
flows, and the luminosity function of quasars,
e.g. \citep{Silk98,Loeb04,Springel05,Haiman06, Novak11,Soker11,Choi14,Nims15,Ciotti16}.
These outflows often manifest themselves as blueshifted broad absorption lines (BALs) 
in 20-40\% of quasars \citep{Hewett03, Dai08}. To assess whether BAL outflows are an effective
agent of quasar feedback, it is necessary to determine 
their average mass flow rate and associated kinetic luminosity.
Theoretical studies and simulations
suggested that kinetic power of order of only ~1\% of the Eddington
luminosity is deemed sufficient for significant feedback effects on
the host galaxy,
e.g. \citep{Scannapieco04,Hopkins06,Hopkins10}, 
which now becomes the benchmark number for observational comparisons.

The spatial extents of BAL outflows are challenging to measure, which
have been found to span several orders of magnitude (ranging from parsec 
to kilo-parsec scales), though pc vs. kpc controversial results are sometimes 
reported in the literature, e.g. see the comments by \citep{Lucy14}.
The controversy in the outflow radius directly leads to large uncertainty
in the outflow enegetics, rendering it pointless to determine whether 
effective feedback is at work. The most direct method to pin down the
outflow radius is using spatially resolved spectroscopy, especially  
integral field unit (IFU) spectroscopy, e.g. \citep{Barbosa09,Riffel11,
Rupke13,Liu13a,Liu13b,Liu14,Liu15,Wylezalek17, Bae17}, but for high-redshift ($z>2$) BAL quasars,
the realistic approach is deriving the galactocentric distance of
the outflow (R) from ionization parameter ($U=Q/[4\pi R^{2}n_{\rm H}c] $), 
where the hydrogen number density $n_{\rm H}$ can be determined
from the absorption lines of the excited states of ions (e.g. Fe II*, Si II*, 
S IV*).
During the last decade or so, the outflow radii have been measured
for a number of individual quasars using density-sensitive absorption lines
from excited levels, e.g. \citep{Hamann01, Arav08, Arav15,
Chamberlain15}. It should be noted that when the utilized absorption 
lines are broad and blended, and/or when severe saturation (commonly) takes 
place, this approach, relying on photoionization modeling and geometric
assumptions, introduces nontrivial uncertainty.

Complementary to the conventional approaches, here we propose a method 
to estimate the spatial location of the outflow 
for a special class of BAL quasars. As illustrated in Fig. 1, a number
of quasars show a narrow (or even spiky) \lya~emission line at the bottom
of broad and close-to-black \nv~absorption trough, indicating that the \lya~
line emission is leaked from the emission line region, which is otherwise
virtually compleletly obscured by the BAL region/outflow
(the concept of a BAL region and that of a BAL outflow are used 
interchangeably throughout the paper, because the detailed physical 
difference between them is subtle; even if the difference is nontrivial,
their spatial scales are expected to be of the same order of magnitude).
With this physical picture constructed, the size of a BAL region can be
derived by assuming virialized line-emitting gas surrounding a super-massive
black hole with known mass (detailed in Section 4).

In this paper, we present three $z=3$ quasars charaterizing close-to-black \nv~
absorption troughs and spiky leaked \lya~emission lines, and demonstrate how
our method is applied on them. As a result of a comprehensive search in the 
SDSS-III DR12 database (see Section 2 for details), these objects are among 
the most representative examples of this special class of BAL quasars. 
This paper is structured as follows. In \S2, we describe the selection of
the quasars from the SDSS-III database, and the basic properties of their
spectroscopy data. The analysis of their spectra is presented in \S3, and 
the size and energetics of their BAL outflows are measured in \S4 and \S5.
We further discuss the origin of these spiky Lya emission lines in \S6, before
a summary given in \S7. Throughout this work, we adopt a standard $\Lambda${\sc CDM} 
cosmology with $\Omega_{\Lambda}=0.7$ ,$\Omega_{M}=0.3$ and $H_{0}=70$\,\kms\,Mpc$^{-1}$.
\begin{figure*}
\center{}
\includegraphics[height=11.5cm,angle=0]{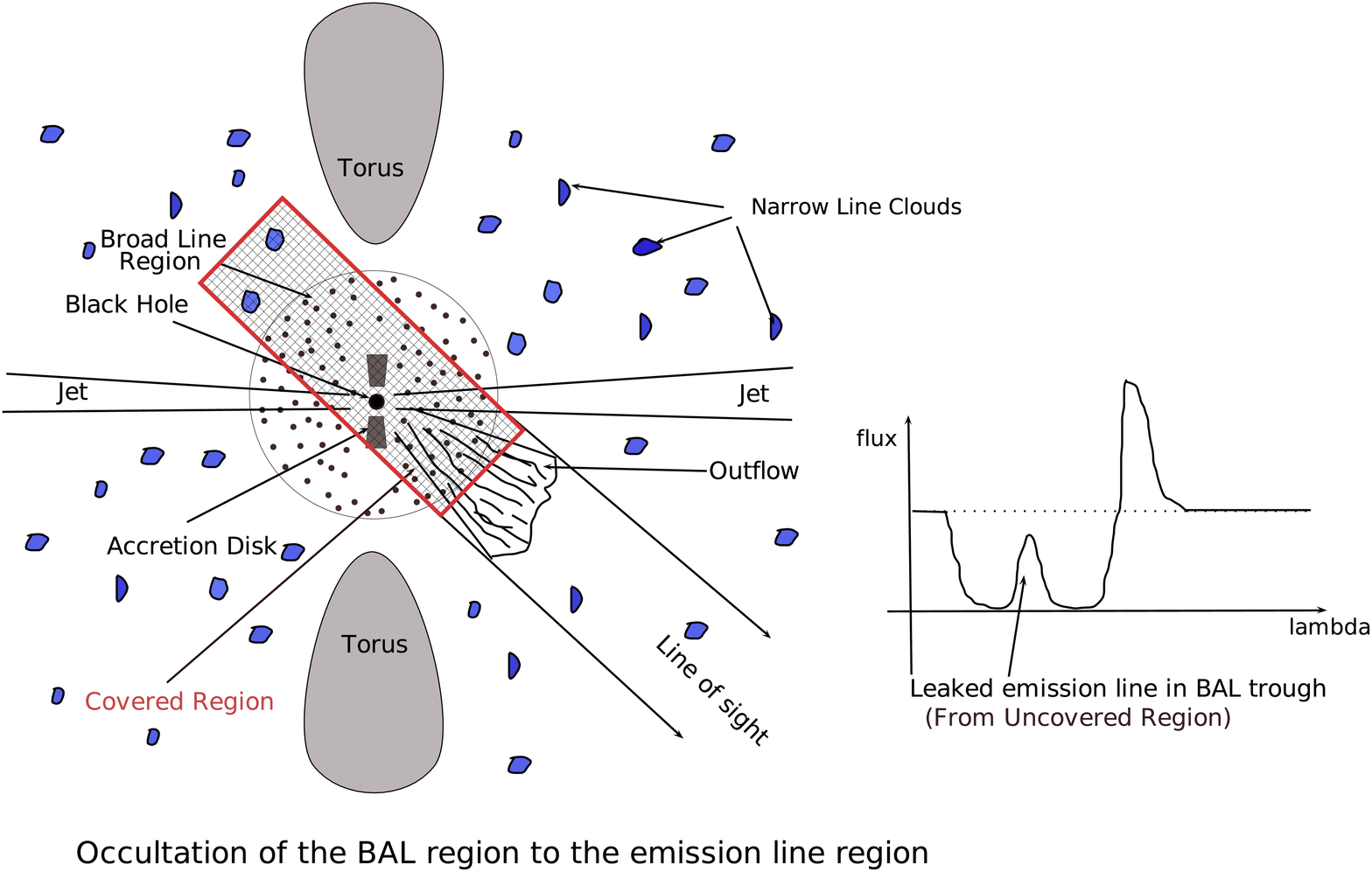}

\caption{ Cartoon of the occultation of the BALR to the emission line region: 
The uncovered part of emission source will show leaked emission lines in the BAL troughs. 
The size of outflow can be estimated by measuring the width of leaked emission line 
by assuming virialized gas in the emission region surrounding a supermassive black hole with known mass. }
\label{fig1}
\end{figure*}
\section{Sample description}
\label{sec:sampleobs}
The Baryon Oscillation Spectroscopic Survey (BOSS), part of the third generation of the Sloan Digital Sky Survey (SDSS-III),
uses the dedicated 2.5-m wide-field telescope at Apache Point Observatory near Sacramento Peak in Southern New Mexico to
conduct an imaging and spectroscopic survey. This catalog is a product of the intensive visual inspection of SDSS
optical spectra from the twelfth data release (DR12; \citep{Alam15}) undertaken
by \citep{Paris16}
\footnote{http://dr12.sdss3.org/datamodel/files/BOSS\_QSO/DR12Q/
DR12Q.html}. About 10\% (29580) of these quasars 
show BALs, whose trough-by-trough properties have been measured and catalogued in the DR12Q\_BAL table
\footnote{http://dr12.sdss3.org/datamodel/files/BOSS\_QSO/DR12Q/
DR12Q\_BAL.html} by the same group.
For our purposes, the visually inspected redshift ($Z_{\rm vi}$), the median signal-to-noise ratio over the rest-frame
wavelength range 1650-1750\AA, and the maximum and minimum velocity that encloses the \civ~absorption trough
(VMAX\_\civ~and VMIN\_\civ, respectively) are the key parameters for our sample selection.

The \LyA~emission line is situated on the blue wing of \NV~emission line with a blueshift velocity of 
5806~$\kms$, and our sample quasars are expected to show the \lya-\nv~region in their SDSS spectra, 
translating to a requirement of $Z_{\rm vi}>2.0$. Among the entire DR12Q\_BAL parent sample, we find 1945
quasars with sufficiently high redshifts.
Meanwhile, \civ~and \nv~are expected to have comparable column densities, because their ionization potentials
(64.5 eV vs. 97.9 eV) and abundances (carbon is only 4.8 times higher) are not
significantly dissimilar. Adopting a criterion of VMAX\_\civ~$> 6500~\kms$
and VMIN\_\civ~$< 5000~\kms$~further reduces the sample size to 290.

Through visual inspection and spectrum fitting, we find 58 quasars whose C IV troughs
show flat bottoms with approximately zero flux. In this paper, we analyze the
3 objects demonstrating the most prominent spiky feature in their Lya emission
line on top of the \nv~absorption trough, and are thus consistent with our
physical interpretion that these lines are leaked out from the BAL absorption
region. The basic information of these sample quasars are reported in Table 1,
with visually inspected redshifts $Z_{\rm vi}$=2.9-3.1, as reported by
\citep{Paris16}.
\section{SPECTRAL ANALYSIS}
\label{sec:Spectral analysis}

\subsection{Redshift calibration}
We observe a systematic blueshift in the \lya, \nv, \siiv~and \civ, when the
visually determined redshift $Z_{\rm vi}$, highlighting the necessity of a line redshift calibration, so that the kinematics of the line-emitting ionized gas can be reliably characterized. Although most of the emission lines (especially the high-ionization broad emission lines) are found to be systematically shifted in the quasar spectra \citep{Grandi82, Wilkes86, Tytler92, Laor95, McIntosh99, VandenBerk01}, an apparent anti-correlation between the velocity shifts and the ionization potential is seen in both broad and narrow lines \citep{VandenBerk01}. From the above anti-correlation, we deduce that the low-ionization \feii~line has an average shift of at the most 385
~\kms, negligible in the composite quasar spectra created by \citep{VandenBerk01}.
Here note that, using \feii~line as redshift reference is new. It can be useful for high-z BAL QSOs, where no other lines in the optical spectrum can be used as a reference.

Employing a single Gaussian profile, we fit the \feii~line (Fig. 2). The resultant calibrated redshifts are 2.909, 3.067, 3.046 for the three quasars (Table. 1).
Applying these redshifts, the centroids of the \lya, \nv, and \civ~lines all show minimal shifts ($<200~\kms$). 
\begin{figure}
\center{}
\includegraphics[height=5.9cm]{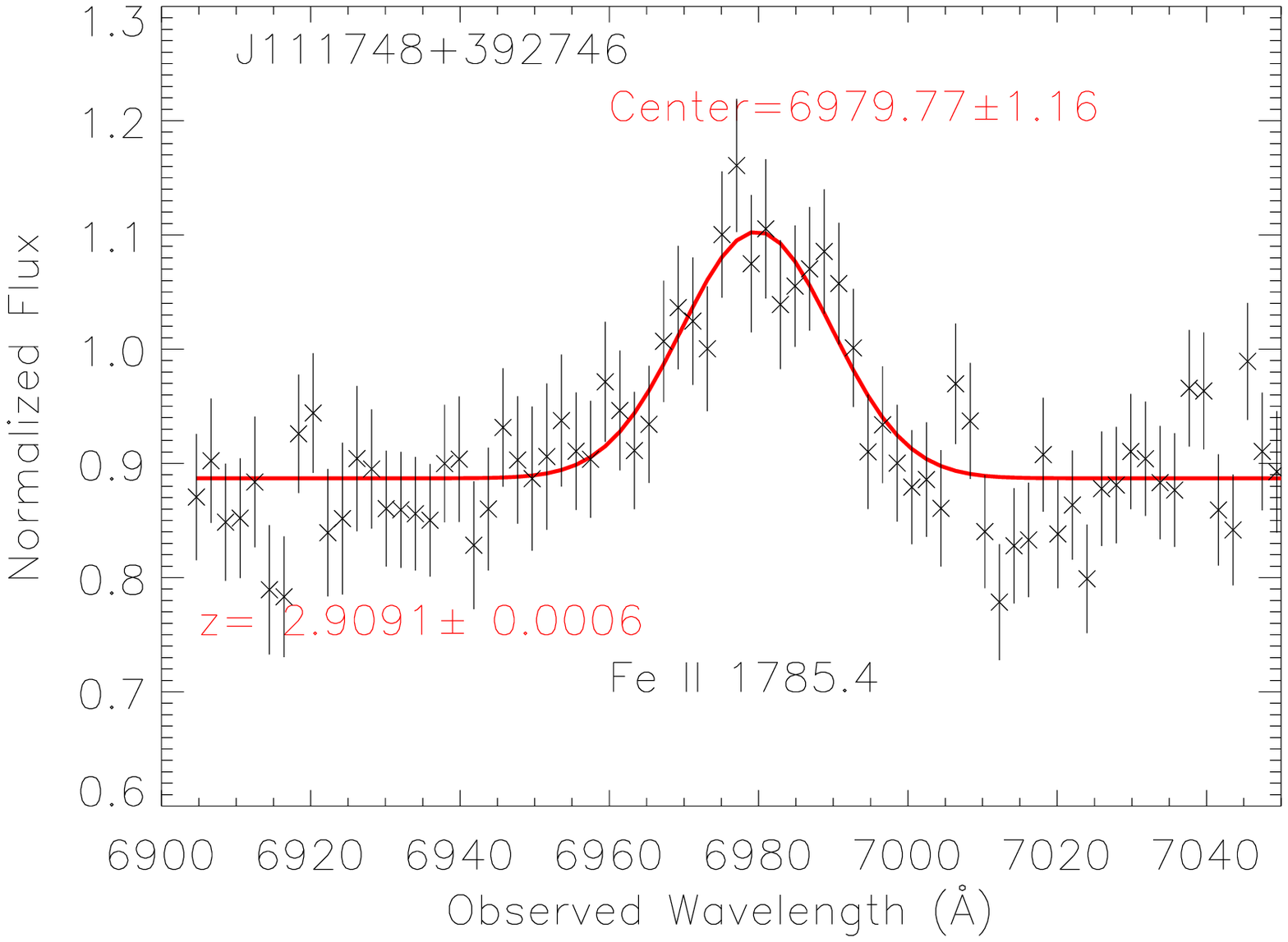}
\includegraphics[height=5.9cm]{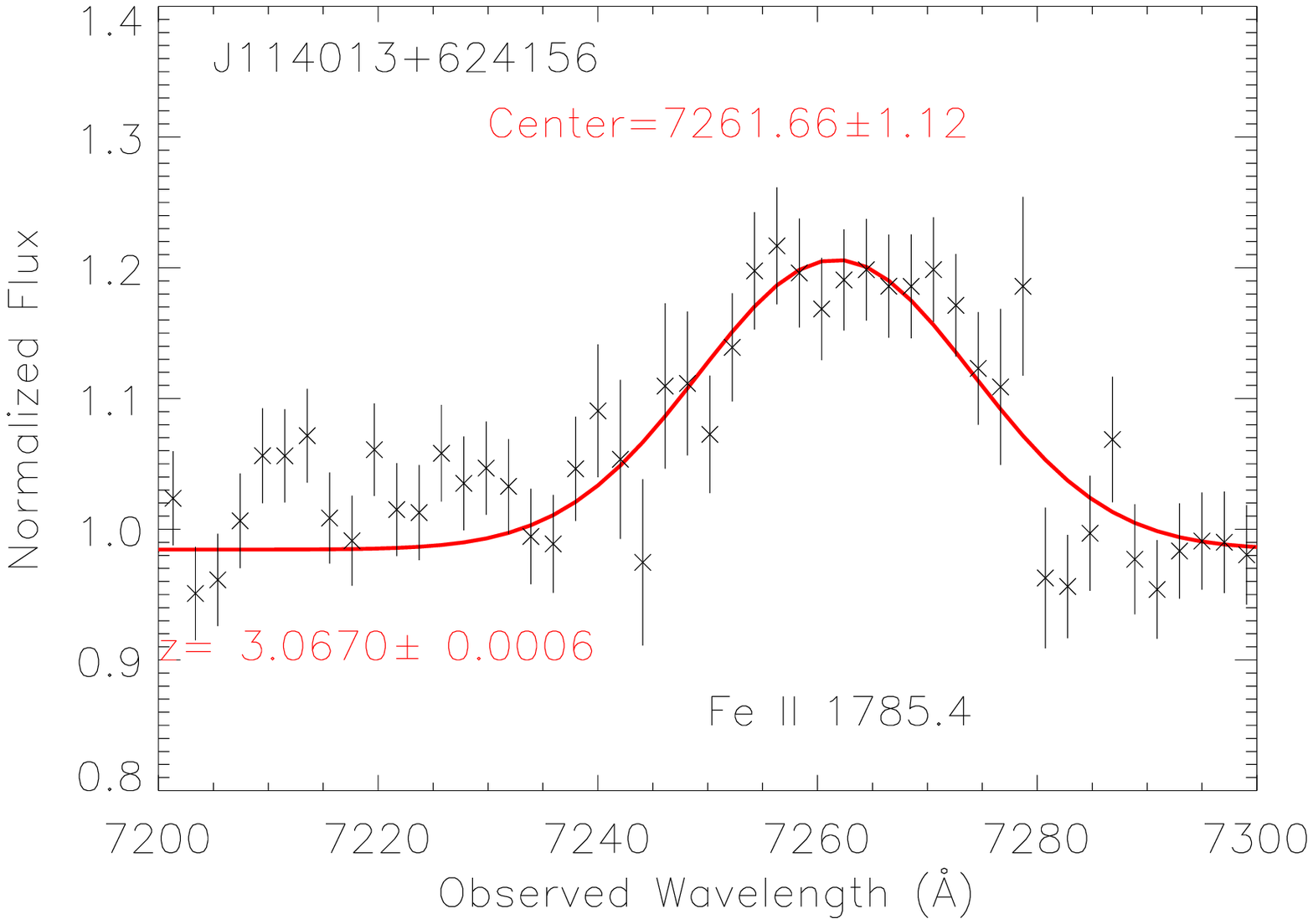}
\includegraphics[height=5.9cm]{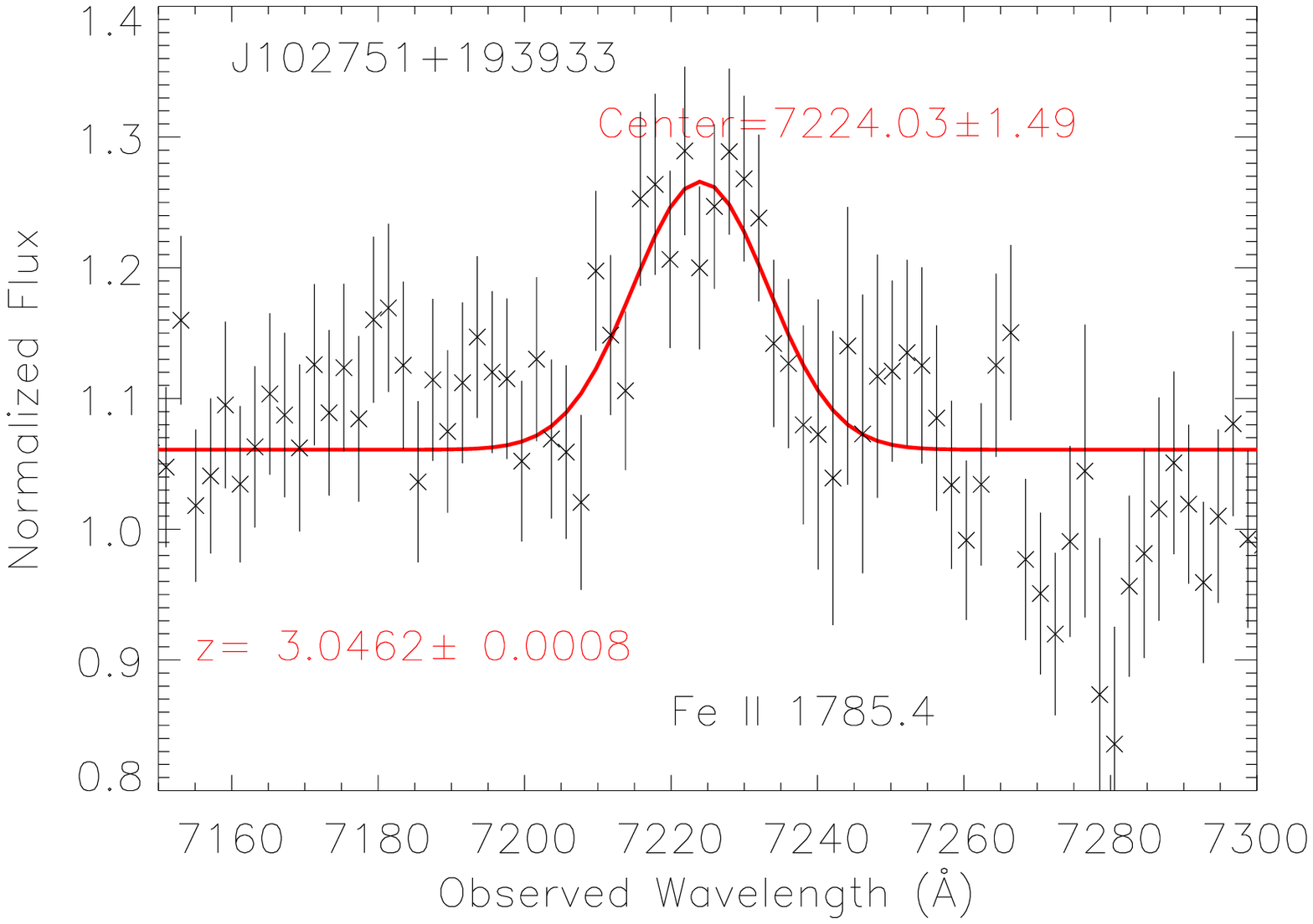}
\caption{Using a single Gauss function to fit the low-ionization \feii~line to give a calibration for the quasar redshift.}

\end{figure}
\subsection{Fitting the spectra}
\label{sec_sub:Fitting the spectra}
The sample quasar spectra all have been corrected for Galactic extinction, adopting the extinction curve of \citep{Cardelli89} (IR band; UV band) and \citep{ODonnell94}(optical band) with $R_{\rm V}$ = 3.1. The $A_{\rm V}$ values of these SDSS BAL quasars are derived from the SpecPhotoAll table in SDSS. To reliably characterize the continuum and delineate it from the \civ, \nv~BAL troughs, we fit the spectra using 165 unabsorbed quasar templates given by \citep{Wang15}, which were drived from 38,377 non-BAL quasars with $1.5 < z < 4.0$ and $SNR_{\rm 1350} > 10$ in SDSS Data Release 7 (DR7). Following \citep{Wang15}, we scale their templates using a scale factor, which is a double power-law
function of the rest-frame wavelength,
\begin{equation}
S_\lambda=A[1]\left(\frac{\lambda}{2000\AA}\right)^{A[2]}+ 
    A[3]\left(\frac{\lambda}{2000\AA}\right)^{A[4]}
\label{eq1}
\end{equation}
where coefficients (A[1],A[3]) and exponents(A[2],A[4]) are pinned down by minimizing the $\chi^2$ value. This fitting spans a wavelength range of 1050 to 2850\AA\, and we add a additional Gaussian component at the \lya, \nv, \siiv~and \civ~locations to the unabsorbed templates to improve the fits therein. Also, we iteratively mask spectral pixels lower than the model with a significance of $\ge 3\sigma$ on the blue side of \lya, \nv, \siiv, \civ~to exclude possible absorption contamination from the \lya~forest. The fitting procedure is demonstrated in Fig. 3.

\begin{figure}
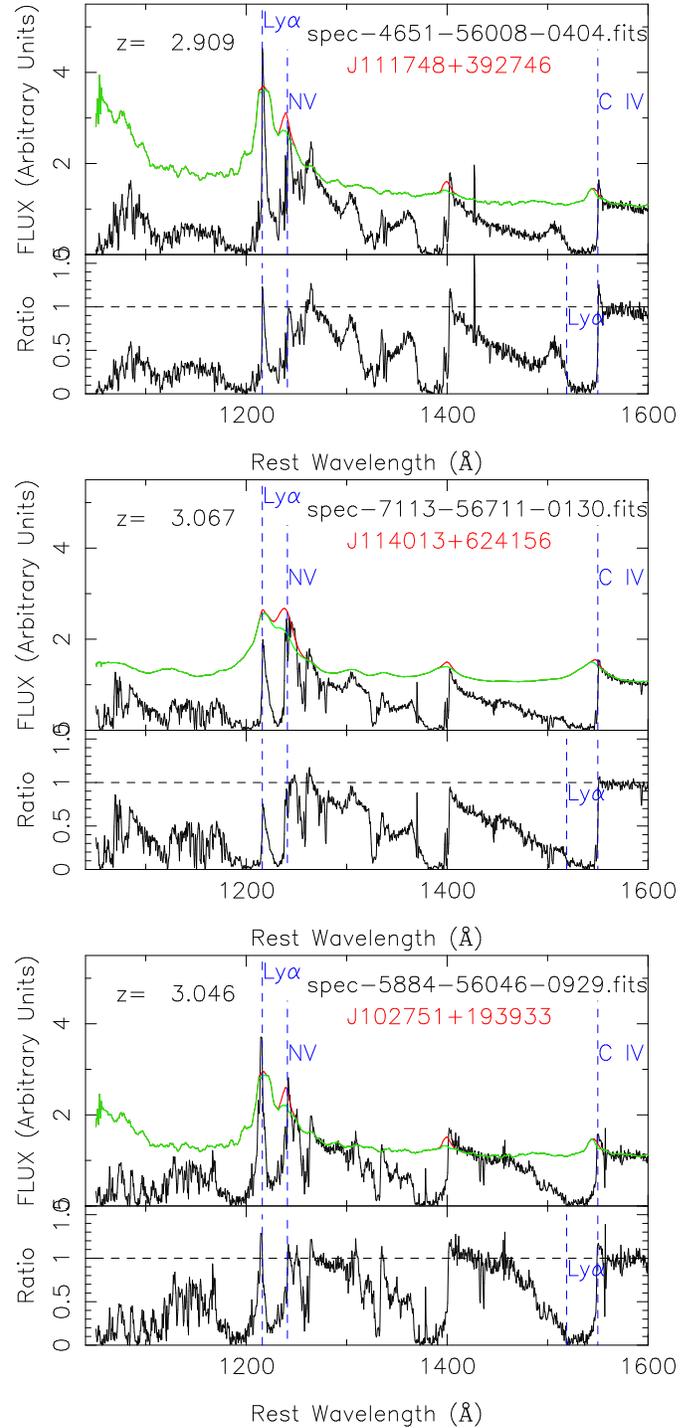

\center{}
\includegraphics[height=8.8cm,angle=-90]{spec-4651-56008-0404.fits.powbest.eps}
\includegraphics[height=8.8cm,angle=-90]{spec-7113-56711-0130.fits.powbest.eps}
\includegraphics[height=8.8cm,angle=-90]{spec-5884-56046-0929.fits.powbest.eps}
\caption{Top panel: unabsorbed template fit to the spectra of the three SDSS quasars.
The cyan line shows the scaled best matched template, while the red line has additional
Gaussians to account for emission lines. Bottom panel:
The ratio of observation to fitting spectrum. The blue dot line mark \LyA~position in
the \nv~BAL trough and the corresponding position in the \civ~trough.}

\end{figure}

\begin{figure}
\center{}
\includegraphics[height=7.0cm,angle=-90]{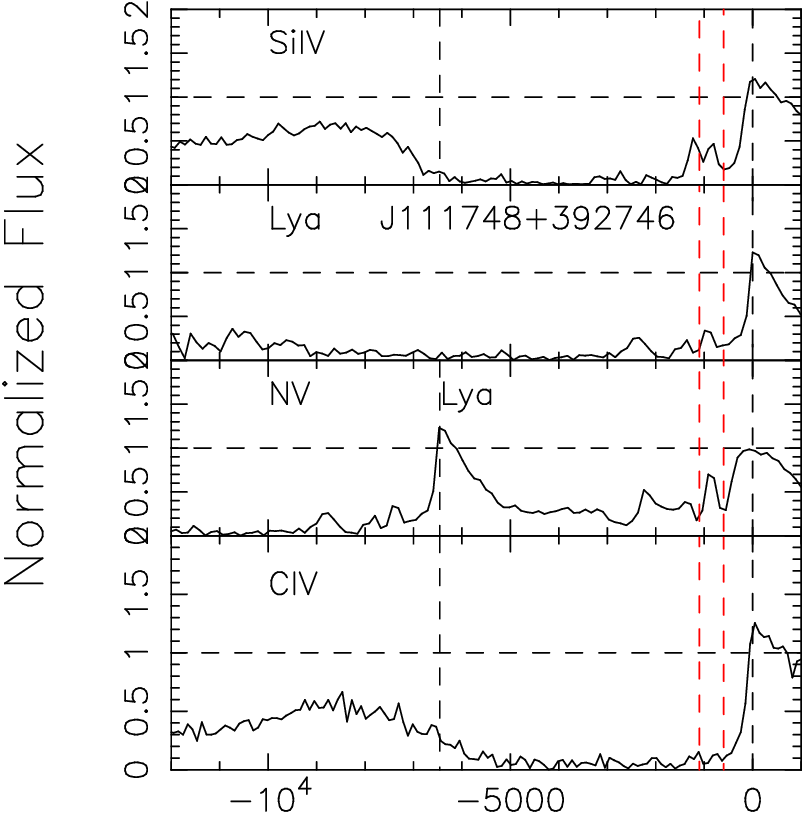}
\includegraphics[height=7.0cm,angle=-90]{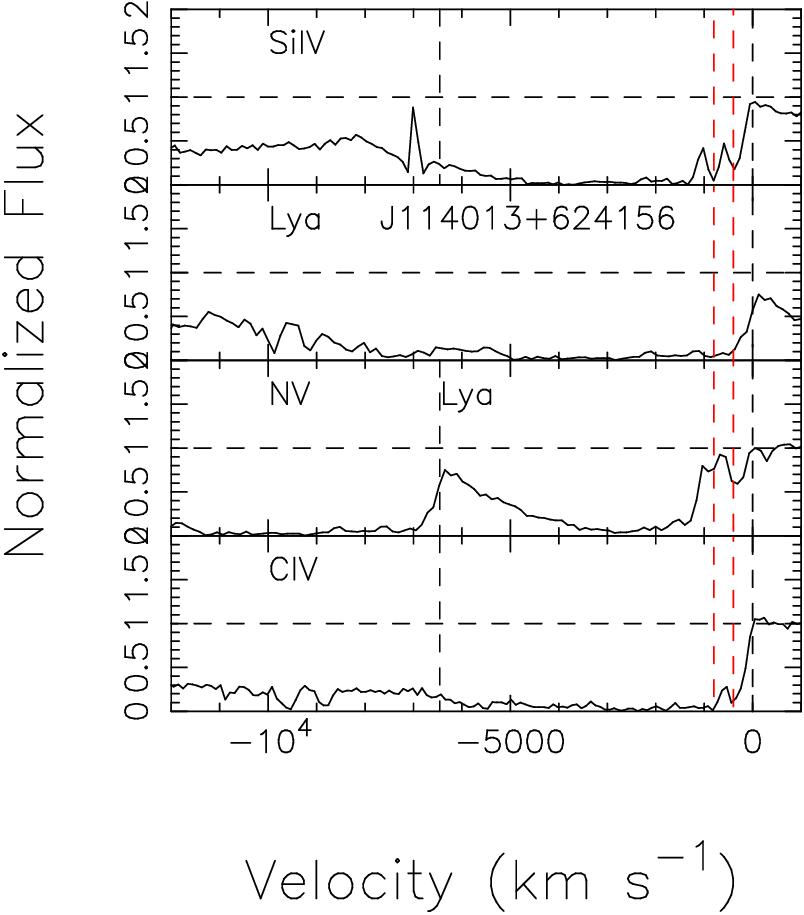}
\caption{Put \civ, \nv, \lya~and \siiv ~ions BAL troughs together in the velocity space. 
The black vertical dotted lines mark the position of red component of other ions or \lya.
The red ones mark the position of the two narrow absorbtions. The velocity of the first narrow troughs relative to the peak of \lya~(or red component of other ion) is about 600~$\kms$ and 400~$\kms$ for J111748+392746 and J114013+624156 respectively. The velocity gap of the two narrow troughs is about 500~$\kms$ for all the ions in both two quasars.}
\end{figure}
Fitting and normalizing the quasar spectra facilitate our characterization of the \nv-\lya~region. As shown in Fig. 3, the bottom of the \nv~trough (in correspondence to \civ) is appoximately flat and the spectral flux is nearly zero (lower than 5\% of the continuum level). In our physical interpretion of the data (see Fig. 1), the \lya~spike is leaked line emission from the unobscured emission region. 
When \civ,\nv(\lya~included) and \siiv~BAL troughs are plotted in the velocity space (Fig. 4), we find two barely separated narrow absorption features at -1000~\kms~to -400~\kms~ in at least two of the four ions in both J1117+3927 and J1140+6241.

\subsection{Characterizing the \lya~emission line}
\subsubsection{\civ~absorption trough}
The residual flux of the \nv~BAL trough likely caused by
partical covering, \citep[e.g.][]{Arav99} or small residuals can be scattered light as 
well because BAL troughs are more polarized, \citep[e.g.][]{Ogle99}.
Although approximating zero, it needs to be fully removed to recover the line profile of the \lya~emission.
For this purpose, we use the spectrum of the \civ~BAL resudial flux as a
template to fit the that of the NV BAL trough. We note that the velocity
separation of the \nv1238.8, 1242.8\AA\ doublet (966~\kms) is larger than that
of the \civ~1548.2, 1550.8\AA\ doublet (503~\kms), and the latter is strongly
blended. 

The complex velocity structure of \civ~prevents us from structuring
a simplestic model (e.g. a single Gaussian), but we have the freedom to
introduce a sophisiticated mathematical formalism, since a phenominalogical
(rather than physical) modeling is sufficient for our purposes. Here, this
formalism is set to be the superposition of a Gaussian on a 5th-order
polynomial. We increase the fitting weight by a factor of 2 for the bottom
of the CIV trough, so that the detailed structure in this region receives
more attention from the fits. As a result, the \civ~velocity profiles are
well represented and de-blended by the best-fit models resulting from this
procedure (Fig. 5).

For the resonance doublet \civ~1548.2, 1550.8\AA\, the ratio of their oscillator
strengths ($f_{\rm blue}$ = 0.19 and $f_{\rm red}$ = 0.095, respectively)
\footnote{http://physics.nist.gov/PhysRefData/ASD/lines\_form.html}, renders
an optical depth ratio of $f_{\rm blue}/f_{\rm red}$ close to 2 (see Equation 9).
Partial covering obscuration is generally assumed to interprete non-black
absorption troughs, for which $I(v)=1-C(v)+C(v)e^{-\tau(v)}$, where $C(v)$ is the covering factor
and $\tau(v)$ is the optical depth of the ion at velocity v \citep{Arav99, Hall03}.
However, for our sample quasars with flat and nearly black troughs, we find it
sufficient to derive the apparent optical depth, which is introduced only as
a proxy for our procedure of analysis. 
Hence, for simplicity and with effectiveness, we take
$C(v)$=1 throughout the absorption trough, and the above formalism retrogrades
to the classical Beer-Lambert law,
\begin{equation}
I(v)=e^{-\tau(v)}.
\label{eq2}
\end{equation}
\subsubsection{\nv~absorption trough}
As shown in Fig. 6, the blue wing of the \lya~emission line
(inside the \nv~BAL trough) is an interplay between \nv~absorption and the intrinsic
\lya~emission (bluer than -6458~\kms, the velocity separation of the \nv~red line
and Lya in our de-redshifted spectrum). To remove the residual flux in the \nv~
trough, we scale the apparent optical depths of the \civ~trough by a factor of $k$,
fit the data to $I_{\rm fit}(v)=e^{-k\tau_{\rm \civ}(v)}$,
and perform the standard $\chi^2$ minimization procedure. The best-fit
k values we obtain are $0.65\pm 0.01, 0.88\pm 0.02$~and$~0.58\pm 0.02$ for SDSS J1117+3927, J1140+6241 and
J1027+1939, respectively. In Fig. 6, the blue line depicts the 'intrinsic profile'
of the \lya~emission as a result of subtracting the best-fit \nv~resudial flux
profile from the observed spectrum.

\subsubsection{\lya~emission line}
We use two methods to measure the width of the leaked \lya~emission line.

(1) Using a single Gaussian to fit the \lya~line (Fig. 7):
we mask the blue wing of \lya~where presents the narrow absorbtion components 
(SDSS J1117+3927 and J1140+6241) when doing the Gaussian fitting.

(2) Directly measuring FWHM of \lya~without fitting (non-parametric).

The results of the two methods are listed in Table 1. Due to the narrow absorptions,
the FWHMs of directly measurement are significantly smaller than that of Gaussian fitting for J1117+3927 and J1140+6241.

\begin{figure*}
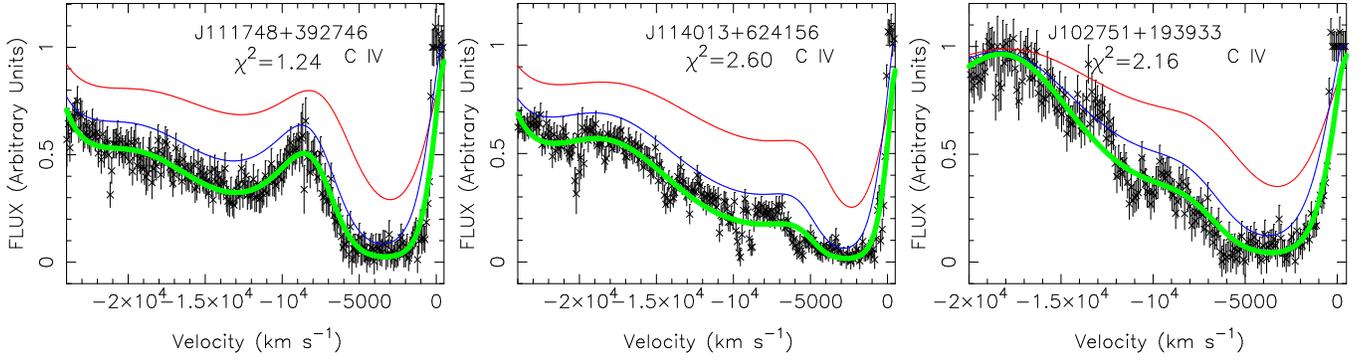

\center{}
\includegraphics[height=5.9cm,angle=-90]{spec-4651-56008-0404.CIV.eps}
\includegraphics[height=5.9cm,angle=-90]{spec-7113-56711-0130.CIV.eps}
\includegraphics[height=5.9cm,angle=-90]{spec-5884-56046-0929.CIV.eps}
\caption{Using a Gaussian plus a five order polynomial function for the optical depth to fit the \civ 1548.2, 
1550.8\AA\ doublet blended BAL trough. Red and blue line repesent the red, blue component of the doublet while 
the cyan one is for the total trough. We use a covering factor c=1 throughout the trough. 
The minimizing $\chi^2$ of the fitting are 1.24, 2.60, 2.16 for J1117+3927, J1140+6241 and
J1027+1939, respectively.}
\end{figure*}

\begin{figure*}
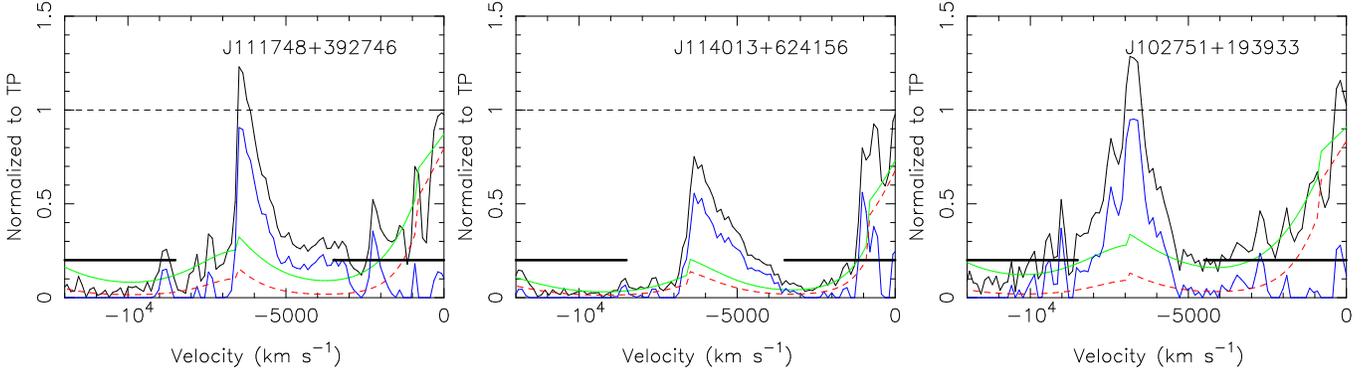

\center{}
\includegraphics[height=5.9cm,angle=-90]{spec-4651-56008-0404.Nv_tao.eps}
\includegraphics[height=5.9cm,angle=-90]{spec-7113-56711-0130.Nv_tao.eps}
\includegraphics[height=5.9cm,angle=-90]{spec-5884-56046-0929.Nv_tao.eps}
\caption{The fitting of the resudial flux of \nv~trough (cyan solid
line) using the extracted template of \civ~ BAL trough (red dotted line). 
The black solid line is the original normalized resudial flux and the blue one is 
the intrinsic (subtracted) normalized resudial flux. 
The fitting window is selected to be away from the leaked \lya~(black horizontal line).  
}
\end{figure*}

\begin{figure*}
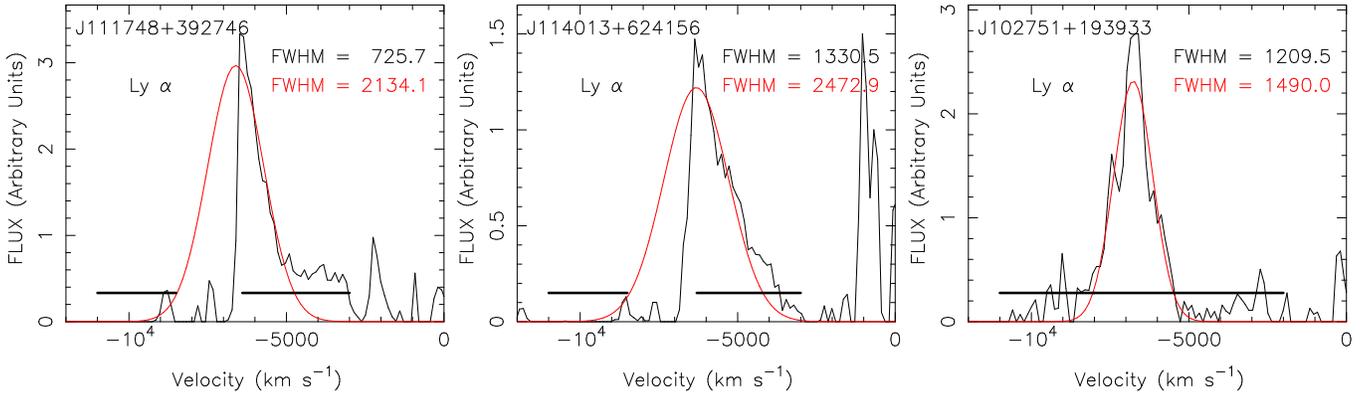

\center{}
\includegraphics[height=5.9cm,angle=-90]{spec-4651-56008-0404.Nvgau.eps}
\includegraphics[height=5.9cm,angle=-90]{spec-7113-56711-0130.Nvgau.eps}
\includegraphics[height=5.9cm,angle=-90]{spec-5884-56046-0929.Nvgau.eps}
\caption{Two methods to measure the width of the leaked \lya~emission line: 
(1) Using a single Gaussian to fit the leaked \lya~line (red).
(2) Directly measuring FWHM of \lya~without fitting (black). 
We mask the blue wing of \lya~where presents the narrow absorbtion components 
(SDSS J111748+392746 and J114013+624156) when doing the Gaussian fitting. The black horizontal line shows
fitting windows.}
\end{figure*}

\section{Results}
\subsection{Methodology}
\label{sec:size}

The radii of the BAL regions (BALRs) can be estimated from the width of the the leaked
\lya~emission. Assumption that the line-emitting gas in this region is fully varialized,
the radius of the \lya~emission line region (i.e. the BALR) is related to the line width
through
\begin{equation}
R_{\rm BALR} =\frac{G\mbh}{fv^{2}}
\label{eq3}
\end{equation}
where the velocity dispersion of the virialized gas $v=\fwhm/(2\sqrt{2\ln{2}})$,
G is $ 6.67\times 10^{-8}cm^{3}g^{-2}s^{-2} $ 
(gravitational constant). $f$ is a virial factor that depends on the geometry and dynamics e.g.,
\citep{Kashi13, Waters16}, with the average values $\langle f \rangle \approx 1-6$ by observations 
\citep{McLure04, Onken04, Woo10, Grier13}. 
Here we adopt $f=4.31$ measured from a sample containing 30 AGNs e.g. see \citep{Grier13}.
Fiducial values for a typical black hole mass and a typical \lya~line width are $10^9\msun$
and $3000~\kms$, respectively, rendering a distance to the central black hole of $\sim$~0.61 pc.
The BALR sizes of the three sample quasars can then be estimated as follows:
\begin{equation}
R_{\rm BALR} =\frac{\mbh}{10^9\msun}\times (\frac{3000~\kms}{\fwhm})^{2}\times  0.61~pc.
\label{eq3}
\end{equation}

\subsection{$M_{\rm BH}$ and outflow sizes}

For our quasars, the SDSS spectrum covers the rest-frame wavelength range of 1100-2500\AA,
and the conventional approach of estimating the bolometric luminosity using the 5100\AA~
luminosity is inapplicable. However, the optical-UV spectrum of a quasar can be
represented by a power law, $ L_{v} \sim v^\alpha$, where the power index alpha is
roughly between 0 and -1 (e.g. Natali et al. 1998). Adopting $\alpha=-1$ results in a constant
lambda $L_{\lambda}$ value across the UV-optical region, and our measured $\lambda L_{\lambda}$ at
2000\AA~directly translates to $\lambda L_{\lambda}$ (5000\AA)=[3.4, 5.4, 2.4]$\times 10^{46}ergs~s^{-1}$,
respectively. Applying a bolometric correction factor of 9 \citep{Kaspi00} on
 $\lambda L_{\lambda}$ (5000\AA), we find their bolometric luminosity to be
[3.1, 4.9, 2.2]$\times 10^{47}ergs~s^{-1}$, respectively (Table 1).

The well-established radius-luminisoity relation allows for deriving the radius of a broad
emission line region $R_{\rm BLR}$ using the following formulism:
\begin{equation}
R_{\rm BLR}=\alpha \left(\frac{\lambda_{\rm 5100} L_{\rm 5100}}{10^{44}ergs~s^{-1}}\right)^{\beta}
\label{eq4}
\end{equation}
where the parameters, $\alpha$ and $\beta$ are $30.2\pm 1.4$ and $0.64\pm 0.02$ given in \citep{Greene05}. 
So, the $R_{\rm BLR}$ are $1.06\pm 0.08$ pc, $1.43\pm 0.09$ pc and $0.85\pm 0.09$ pc for the three quasars respectively. 
It should be pointed out that, UV luminosity is better than optical one for R-L relation if there is no extinction 
because it is the ionizing continuum causes emission lines.
Since high-luminosity quasars have $L_{\rm Bol}/L_{\rm Edd} \sim$ 1, their MBH values are about
[2.4, 3.9, 1.7]$\times 10^9\msun$, correspondingly. 
Note that, there are black hole formulaes for UV continuum/line, especially for \MgII~\citep{Wang09} which is not in the wavelength range of our quasar spectrums.
Inserting these numbers into Equation (3),
we find the following outflow radii (Table 1):
 
(1) J1117+3927: 2.96 pc (Gaussian fit) or 25.62 pc (non-parametric);

(2) J1140+6241: 3.50 pc (Gaussian fit) or 12.09 pc (non-parametric);

(3) J1027+1939: 4.26 pc (Gaussian fit) or 6.46 pc (non-parametric).

As mentioned in \S3.3.3, due to the narrow absorptions, the FWHMs of directly measurement 
are significantly smaller than that of Gaussian fitting for J1117+3927 and J1140+6241. 
As a result, the $R_{\rm BLR}$ deduced from FWHM of directly measurement is larger than 
that from Gaussian fitting. We take the $R_{\rm BLR}$ deduced from FWHM of directly 
measurement as an upper limit. These results indicate that sizes of the BAL regions of 
the three sample quasars are roughly two orders of magnitude larger than the theoretically 
predicted sizes of the trough forming region (0.01-0.1 pc) for accretion disc line-driven 
winds \citep{Murray95, Proga00}, but is comparable to those
determined from BAL variability \citep{Capellupo11,Shi16}.
Two outflow components are also found to be between 1 and 10 pc from the central
source \citep{de02a, de02b}. They lie closer to the central source than most
of the other outflow which deduced using troughs from excited states e.g. \citep{Hamann01, Arav08, Arav15,
Chamberlain15}.

\begin{table*}
\centering
\caption{Main physical characteristic parameters of the three BAL quasars. Z is the redshift measured using \feii~line. 
The values in the brackets of $\fwhm$ (\lya) and $R_{\rm BALR}$ are for the directly measuring FWHM of \lya~without
fitting (non-parametric).}

\begin{tabular}{lcccccccccccccccc}
\hline
  NAME   &  $Z$ &  \lbol & \mbh  & FWHM (\lya) & $R_{\rm BALR}$ & $R_{\rm BLR}$     \\
         &     &$ (10^{47}ergs~s^{-1})$        &$(10^9\msun)$      & (\kms) & (pc)        & (pc)    \\
\hline
J111748+392746 &2.909       & $3.08\pm0.22 $         &2.44        & 2134.1(725.7) & 2.96(25.62)     & 1.06$\pm 0.08$       \\
J114013+624156 &3.067       & $4.88 \pm0.20 $         &  3.87     &2472.9(1330.5) & 3.50(12.09)      & 1.43$\pm 0.09 $     \\
J102751+193933 & 3.046      & $2.16 \pm0.22  $         & 1.71     &1490.0(1209.5) & 4.26(6.46)    & 0.85$\pm 0.09 $      \\

\hline
\label{t1}
\end{tabular}
\end{table*}

\begin{table*}
\centering
\caption{Main physical parameters of the BAL outflow in the three BAL quasars.}

\begin{tabular}{lcccccccccccccccc}
\hline
  NAME  &  $V_{\rm centroid}$~(\civ) & $N_{\rm \civ}~$ & $N_{\rm \siiv}~$ & $N_{\rm \aliii}~$ & $N_{\rm H}~$ & $\dot{M}$ & $\dot{E_{\rm k}}/\ledd$   \\
   &$(\kms) $ & $(\cmii) $  & $(\cmii)$  & $(\cmii)$  & $(\cmii)$   &  $(\mpyr)$  & (\%)     \\
\hline

J111748+392746 &11543    & $\ge 7.62E+16 $  & $\ge 1.28E+16$  & $1.09E+15$  & $\ge 3.55E+21$   &  $\ge 3.45$  &$\ge 0.05$       \\
J114013+624156 &11526    & $\ge 3.87E+16 $ & $\ge 1.08E+16$  & $6.65E+14$ &  $\ge 1.58E+21$    & $\ge 1.75$   & $\ge 0.02$     \\
J102751+193933 &7698     & $\ge 2.10E+16 $ & $\ge 1.10E+16$  & $1.76E+15$ &  $\ge 9.12E+21$    &  $\ge 8.68$  & $\ge 0.07$      \\

\hline
\label{t2}
\end{tabular}
\end{table*}
\section{Outflow energetics} 
\label{sec:energetics}
\begin{figure}
\center{}
\includegraphics[height=5.5cm]{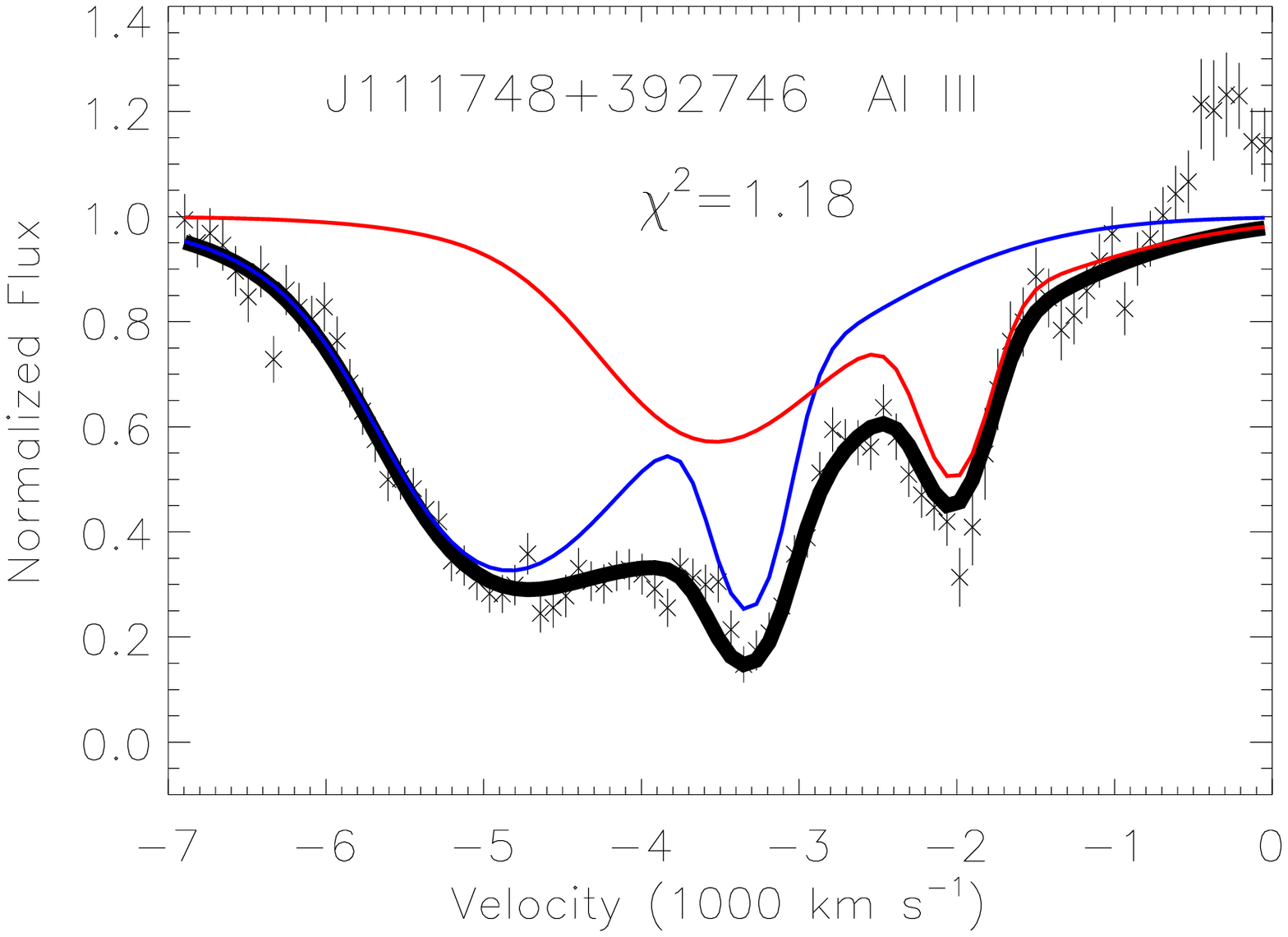}
\includegraphics[height=5.5cm]{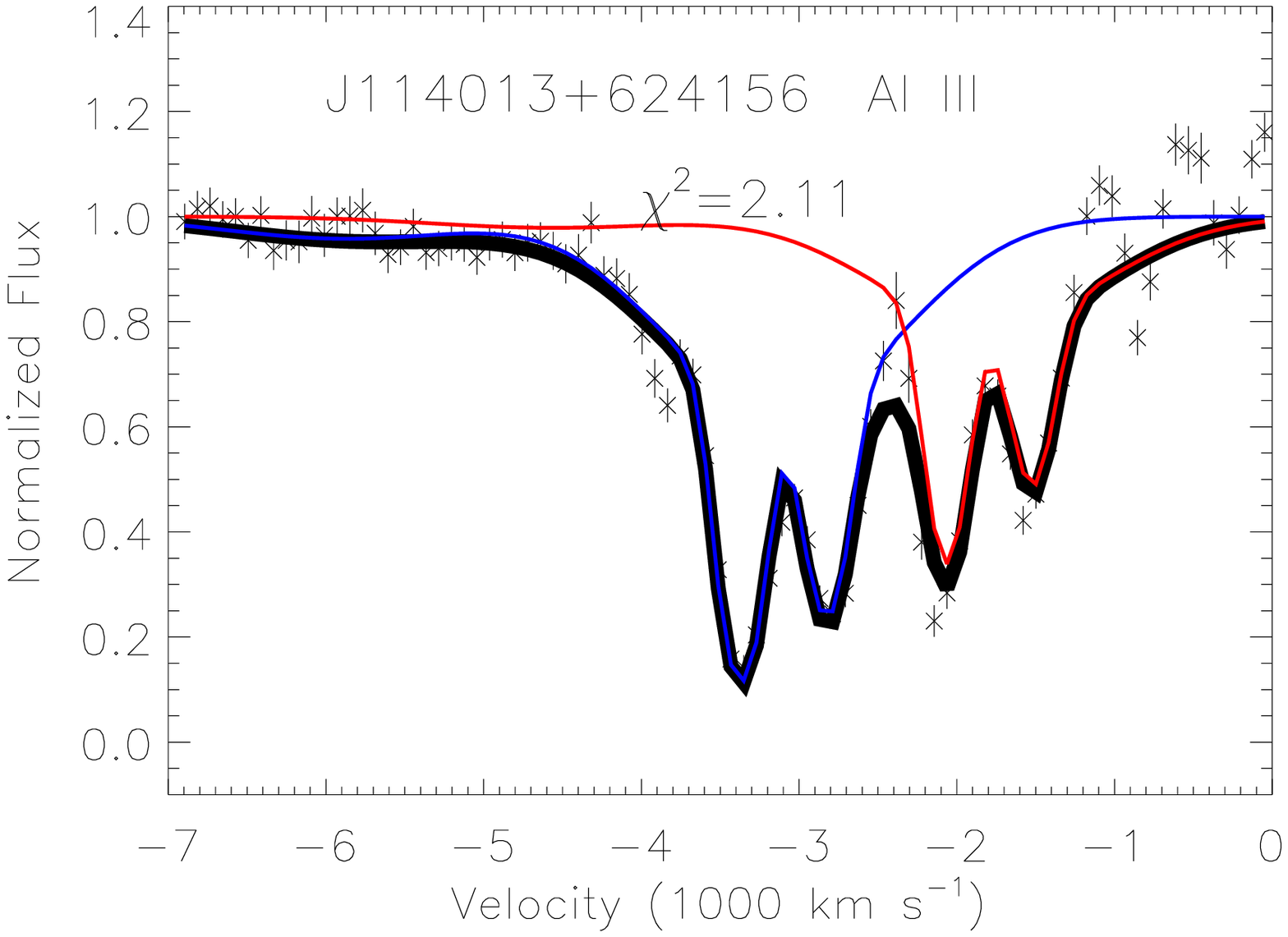}
\includegraphics[height=5.5cm]{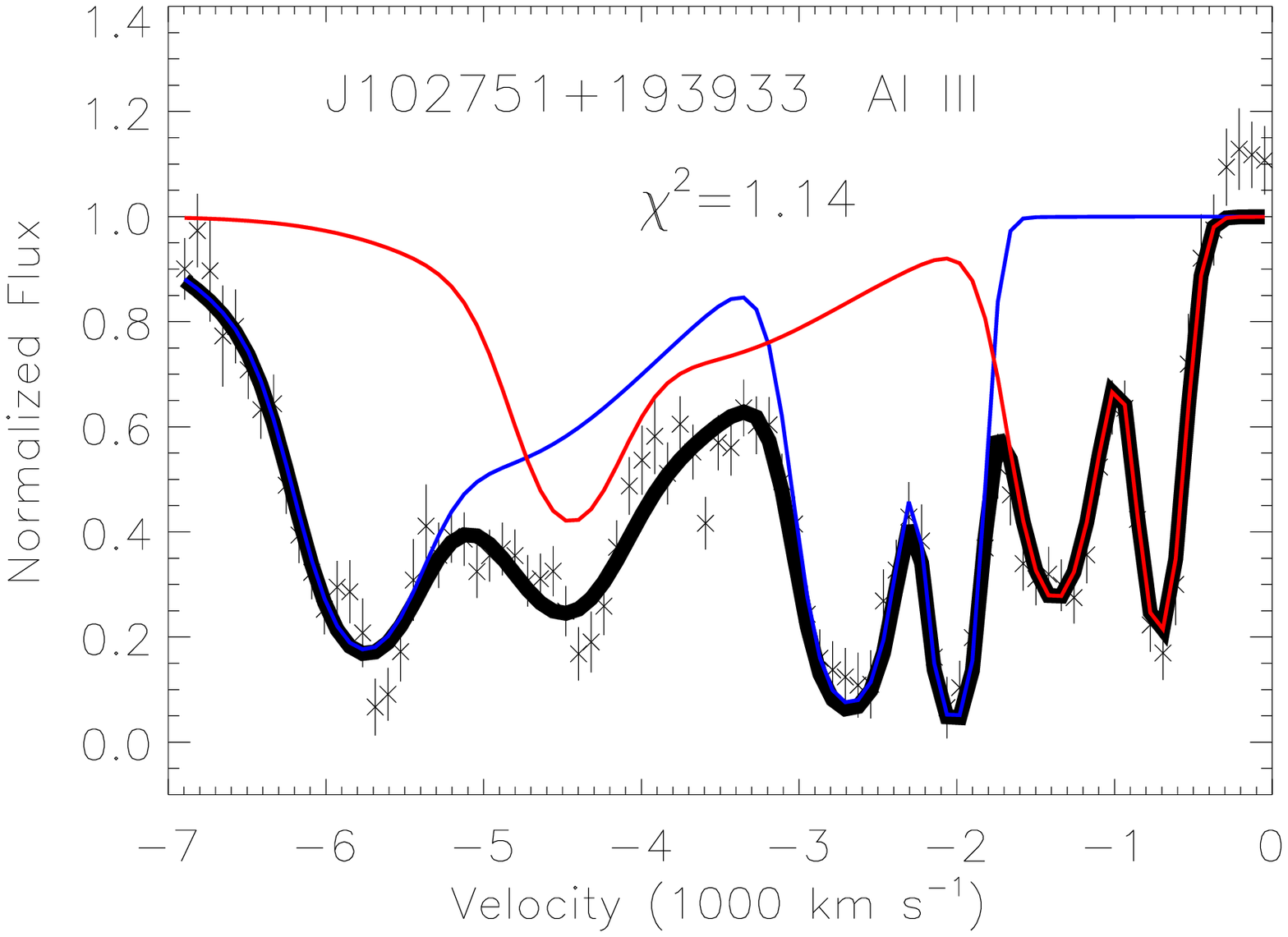}
\caption{Using two to four Gaussian functions as the form of optical depth of the \aliii1854.7, 1862.8\AA~doublet blended BAL trough.
Considering the flat and deep absorbtion of \civ~and \siiv~trough, we take the covering factor c=1 throughout the trough when doing the fit.
The red line repesents the red component while blue repesents the blue one.
The minimizing $\chi^2$ of the fitting are 1.18, 2.11, 1.14 for J1117+3927, J1140+6241 and
J1027+1939, respectively.}
\end{figure}
\begin{figure}
\center{}
\includegraphics[height=5.5cm]{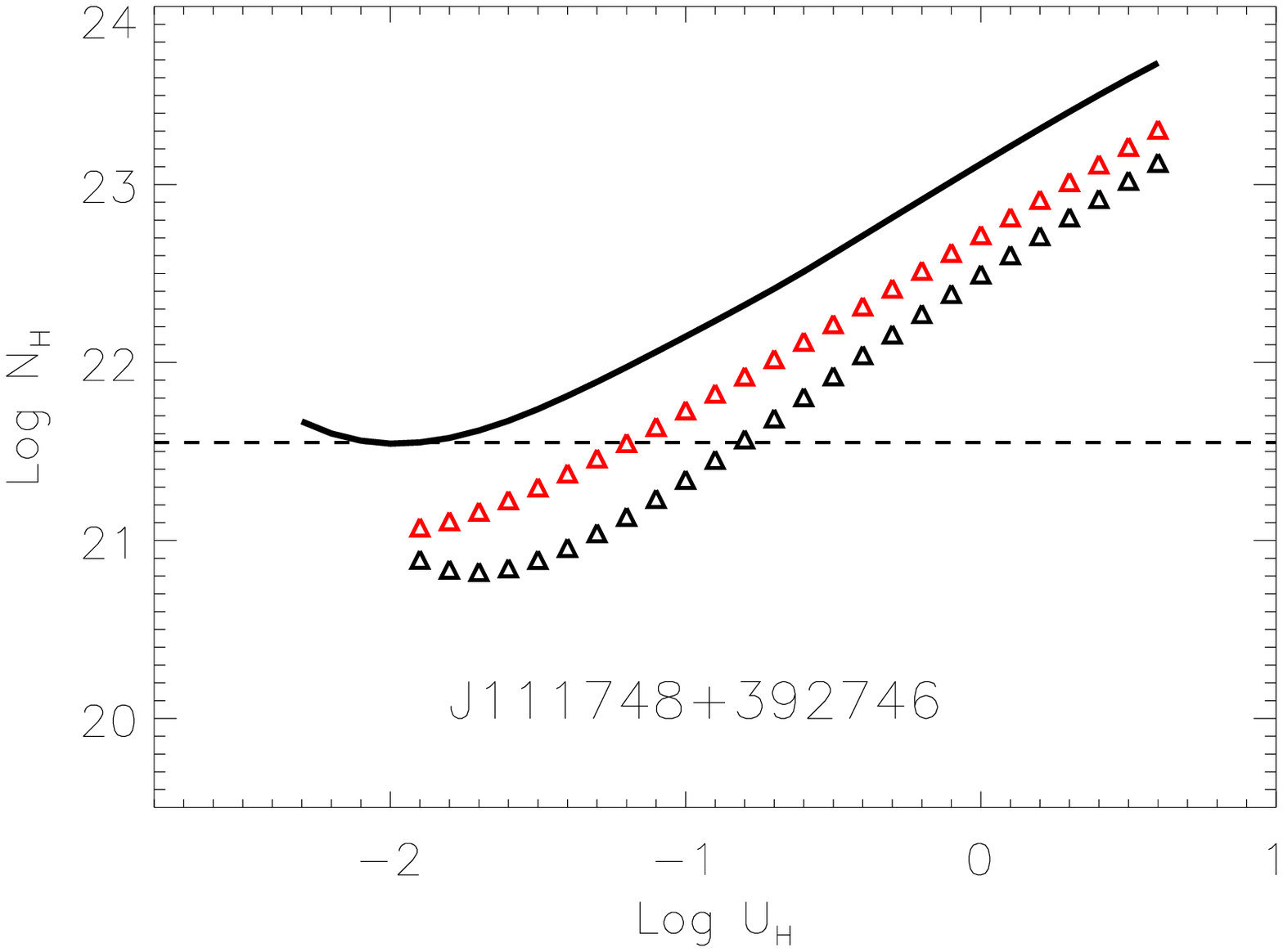}
\includegraphics[height=5.5cm]{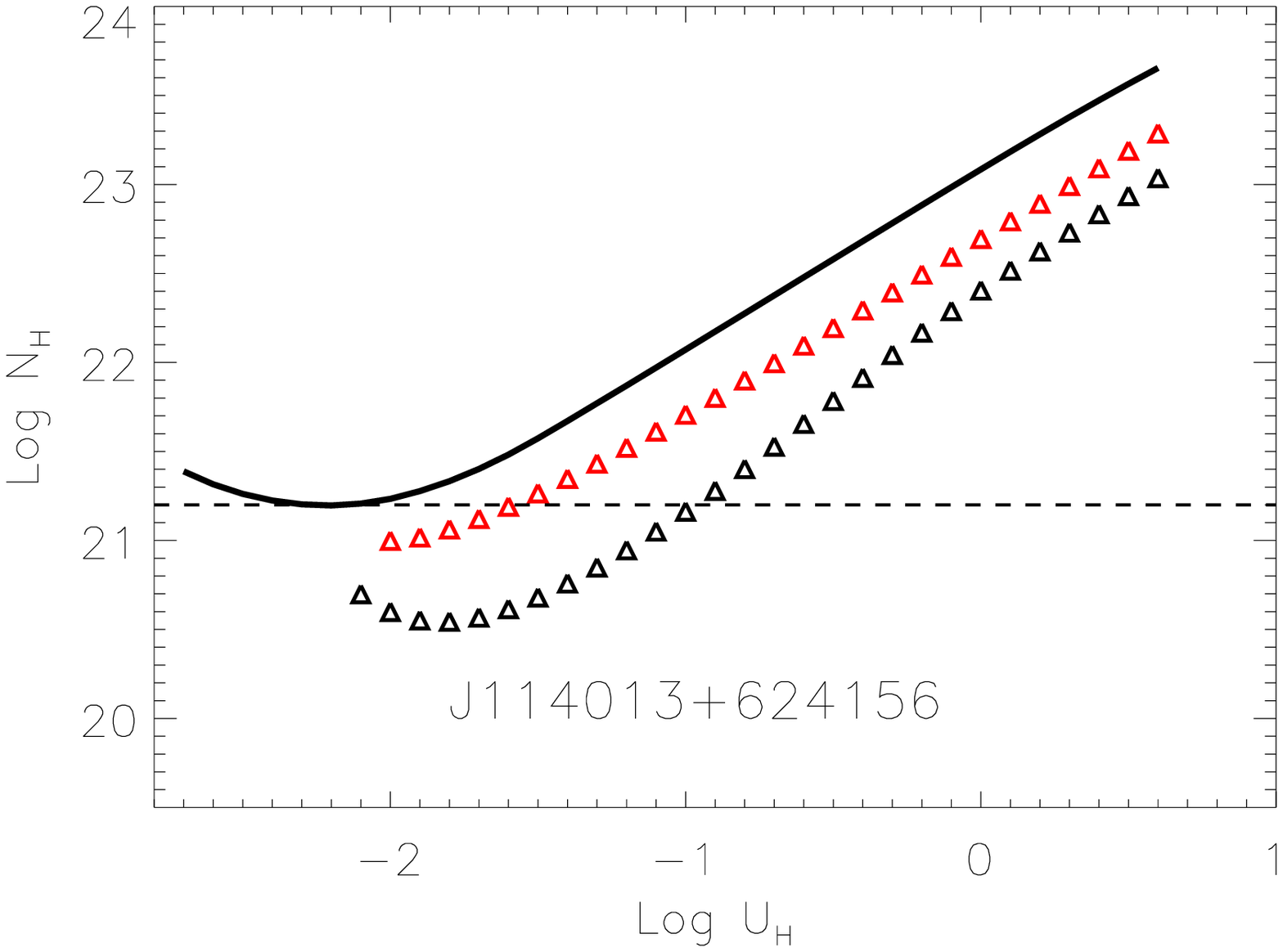}
\includegraphics[height=5.5cm]{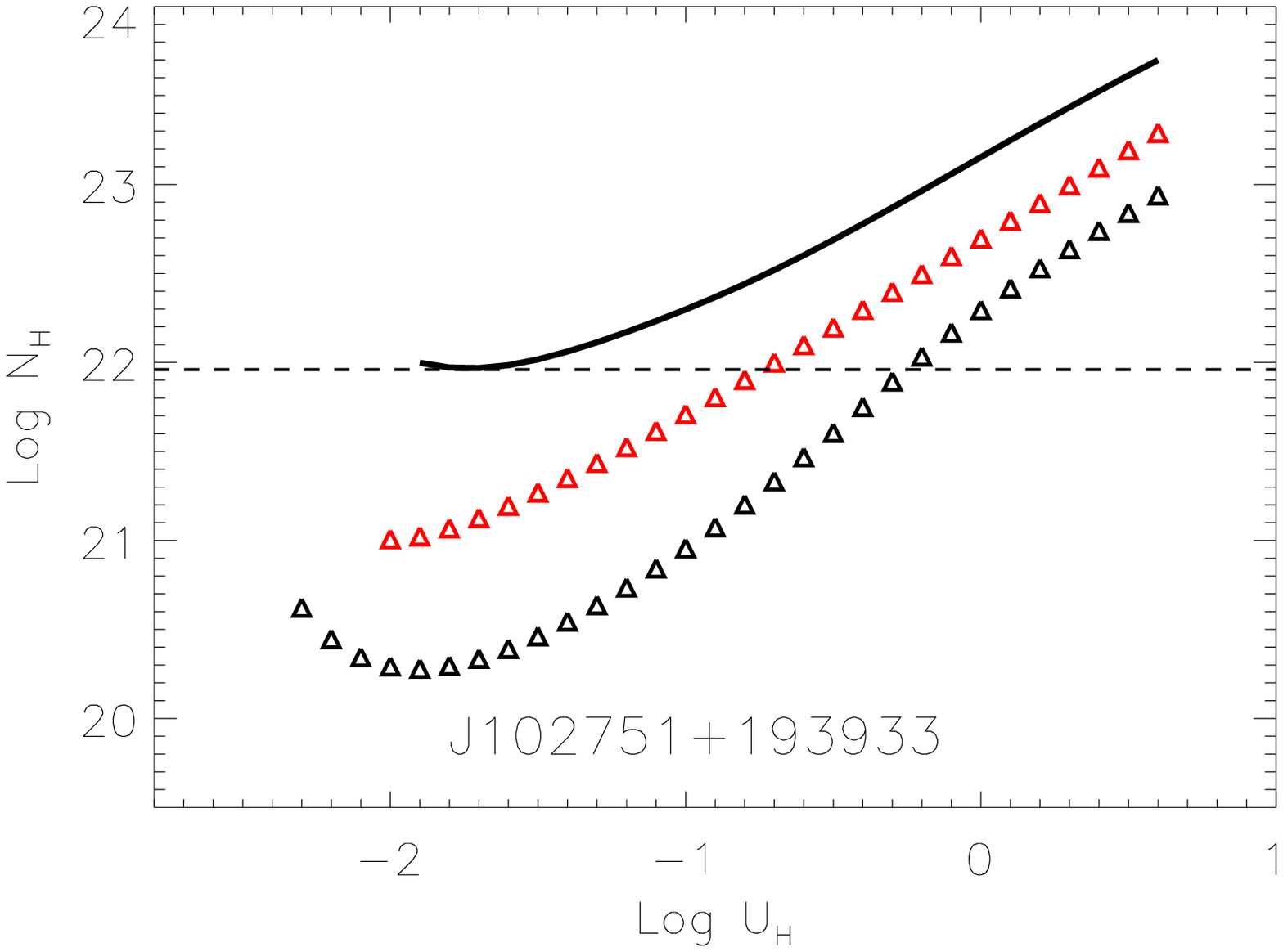}
\caption{$N_{\rm H}-U_{\rm H}$ plane plot showing the photoionization solution using the UV-soft SED for gas with one solar metallicity. The black triangle repesents the lower limit of $N_{\rm H}$ from $N_{\rm \civ}$ while red triangle from $N_{\rm \siiv}$. 
The black line represents the $N_{\rm H}$ deduced from $N_{\rm \aliii}$.
The dotted line repesents the finally lower limit of $N_{\rm H}$ which used to calculate the lower limit of the mass-flow rate $\dot{M}$ and kinetic luminosity $\dot{E_{\rm k}}$ of the outflow.}
\end{figure}

Quantifying the effectiveness of AGN feedback requires an estimate of the mass flow
rate $\dot M$ and kinetic luminosity $\dot E_{k}$ of the outflow. Adopting the conventional
assumption of a partial thin shell geometry ($\Delta R/R\ll 1$), these quantities are given
by \citep{Borguet12},
\begin{equation}
\dot{M} =4\pi\Omega R\mu m_{p} N_{\rm H}v
\label{eq5}
\end{equation}
\begin{equation}
\dot{E_{k}} =2\pi\Omega R\mu m_{p} N_{\rm H}v^3
\label{eq6}
\end{equation}
, where R is the distance from the outflow to the central source,
$\Omega$ is the global covering fraction of the outflow, $\mu$= 1.4 is the mean atomic mass per proton, 
$m_{p}$ is the mass of the proton, $N_{\rm H}$ is the total hydrogen column density of the absorber, 
and $v$ is the radial velocity of the outflow.
Here, we adopt the weighted centroid velocity of \civ~BAL trough, i.e.,
the mean of the velocities where each data point is weighted with its distance from the normalized 
continuum level \citep{Filiz13}.
From the Equation (3) and (7), getting the 
\begin{equation}
\begin{split}
\dot{E_{\rm k}}/\ledd
&=\frac{\Omega}{0.2}\times (\frac{v}{10^{4}~\kms})^{3}\times \frac{N_{\rm H}}{10^{21}\cmii}\\
&\times (\frac{3000~\kms}{\fwhm})^{2}\times 0.004323\%.
\label{eq6}
\end{split}
\end{equation}
Although we derived the black hole mass in the last section, it should be 
noted that $M_{\rm BH}$ does not appear in Equation (8), so the accuracy of estimating
$M_{\rm BH}$ does not affect the characterization of the outflow energetics. The
fiducial number $\Omega$ $\sim$ 0.2 is derived from the convention that the fraction 
of BAL quasars among all quasars (defined from the \civ~absorption trough) 
statistically represents the average fraction of $4\pi$ covered by the solid 
angle of the outflow (see Introduction). The total hydrogen column density 
$N_{\rm H}$ could be deduced from the observed column density of \civ, \siiv~and \aliii~
through photoionization modeling(e.g. CLOUDY; \citep{Ferland98}).

We use a combination of 2-4 Gaussians to fit the optical depth profile of the 
BAL trough as a result of the blended \aliii1854.7, 1862.8\AA. As we did for
the \civ~trough, we adopt a covering factor of C=1 thoughout the entire \aliii~
trough when it is fitted to the Gaussians. 
Integrating the optical depth over
the trough yields the ionic column density \citep{Savage91}:
\begin{equation}
N_{ion} =\frac{3.7679\times10^{14}\cmii}{\lambda f}\int{\tau(v)dv}
\label{eq7}
\end{equation}
where $\lambda$ and $f$ are the wavelength and oscillator strength of the 
transition, and the velocity is the unit of \kms. It should be noted
that the above phenomelogical calculation likely gives a lower limit
of the \civ~and \siiv~column densities, regarding the broad and close-to-black 
absorption that is likely saturated. The results are summarized in Table 2.

In order to further derive the hydrogen column density of the absorbers
(again, actually a lower limit) from that of \civ, \siiv~and \aliii,
we perform a series of photoionization simulations using version c13.03 
of CLOUDY \citep{Ferland13}. We adopt a typical density $n_{e}$ = $10^6\cmiii$,
considering that gas ionization is not sensitive to electron density at a 
given ionization parameter U e.g. \citep{Wang15}. We compute a set of models over the range of
ionization parameter $-3\le Log U\le 1$ using a step of $\Delta Log U$ = 0.1. 
Due to the fact that the three sample quasars in this paper are all 
high-luminosity, radio-quiet quasars, we use the UV-soft SED for this
quasar type following \citep{Dunn10} to characterize the continuum
of ionizing photons. We assume the solar metalicity for these quasars,
though previous discussion in the literature has shown insiginificant
effects when the metalicity is varied in a reasonable range \citep{Chamberlain15}.

The results are plotted on the $N_{\rm H}$-$U_{\rm H}$ plane in Fig. 9. The black triangles
depict the lower limit of $N_{\rm H}$ derived from $N_{\rm \civ}$, 
while red triangles are those from $N_{\rm \siiv}$. 
We adopt the lower limit of $N_{\rm H}$ derived from \aliii~(the dotted line),
which is used to calculate and constrain $\dot{M}$
and $\dot{E_{\rm k}}$ of the outflow. Correspondingly,
the lower limit of $\dot{M}$ is 1.4, 1.0 and 6.1 \mpyr for
J1117+3927, J1140+6241 and J1027+1939, and the lower 
limit of $ \dot{E_{\rm k}}/\ledd$ is 0.05\%, 0.02\%, 0.07\%, respectively.
It is worth mentioning that \citep{Kurosawa09} predicted the quasar outflow 
efficiencies are as low as found in our work.

\section{Discussion: the origin of spiky \lya~emission lines}
\label{sec:disscusion}
In this paper, we interpret the spiky \lya~emission lines on top of flat,
nearly black \nv~BAL troughs to be leaked emission from the broad line
region. The line width of the \lya~lines is $\sim$~2000~\kms~(comparable to
that of broad emission lines), which, in our physical picture, orginates 
from the roughly virialized line-emitting gas. In this section, we discuss
on other possible origins of these \lya~spikes, and assess their pausibility.

(1) Galactic disks:
the \lya~line width of our sample quasars are all over $10^3$~\kms,
which safely rules out the possibility that the spiky \lya~originates
from galactic disks, for which the ionized gas produced by star formation
activity generally has a velocity dispersion of the order $< \sim 100-200~\kms$ 
e.g. \citep{Liu13b}. 

(2) Ambient gas/outflow: 
an outflow that extends on galactic (or even intergalactic) scales may 
produce a pair of superbubbles, as is seen in a number of quasars 
e.g. \citep{Greene12,Liu13b}. 
The wall of the superbubbles is generally thin, and its relatively low 
velocity dispersion may cause a spiky component in the \lya~emission line.
In fact, both recombination and resonant scattering may produce \lya~ 
photons, but the large column density of the BAL troughs in the three 
sample quasars indicates their outflows to be highly optically thick to 
the hydrogen-ionizing continuum. In that case, the chance that the 
ionizing continuum photons escape to (inter-)galactic scales in
the directions of BAL outflows (i.e. our line of sight) and produces
the recombined or scattered \lya~emission is expected to be minimal.
Admittedly, the possibility cannot be completely ruled out that the 
spectrum also collects \lya~light from galaxy-wide superbubbles outside 
the (strictly defined) line of sight, but the rather low surface 
brightness feature generally found in superbubbles does not appear
to be consistent with the strong \lya~emission seen in our sample.
In addition, the line can be resonantly scattered line by a rotating 
outflow \citep{Wang07}, in this case, the line width is of order 
the rotation velocity, which can be either sub or super-Keplerian depends 
where there are strong magnetic fields.

(3) Intergalactic gas in cold accretion: 
numerical simulations predict that the rapid replenishment of gas for 
star formation may have proceeded in the "cold accretion" mode 
\citep{Keres05, Dekel09}. 
The cold gas accreted into galaxies in the cosmic web can be photoionized 
by the quasar, or even the neutral hydrogen therein may scatter the quasar 
radiation to give rise to \lya~emission. The origin of circumgalactic gas
remains unconclusive, but evidence has been found indicating the combination
of pristine gas accreted in this manner and gas reprocessed by AGN and/or 
star formation activity e.g. \citep{Lehner13, Chen16}.
Although a challenging task, the extended \lya~emission from cosmic-web nebulae has been 
detected in $z\sim~2$ to 3 quasars \citep{Cantalupo14, Martin14, Hennawi15}. 
However, the velocity dispersion of the intergalactic gas in cold accretion 
is expected to be $<50~\kms$, smaller than our targets by about 2 orders of 
magnitude. In addition, our searching in the SDSS archive leads to no spotted
quasars within a distance of $\sim$~2.4 Mpc from our targets, while clustered quasars 
hint for the existence of a giant nebula and a proto-cluster \citep{Hennawi15}.

\section{SUMMARY}
In this paper, we present an estimate of the size of BAL outflows for a special
type of quasars, whose \nv~absorption troughs are wide, flat and nearly black,
enclosing a spiky \lya~emissin line. Our interpretation is that the spiky \lya~
emission is leaked from the BAL material obscuring the broad emission line region.

Our systematic search in the SDSS DR12Q catalog renders 3 quasars prominently
characterizing the above feature. Under the assumption that the line-emitting gas
is virialized, we estimate that the FWHM of the \lya~spikes ($\sim$~2000~\kms) 
indicates a size of their outflows to be of the order 3-26 parsecs, similar or
moderately larger than the theoretically investigated trough forming region 
(0.01-0.1 pc) for accretion disc line-driven winds \citep{Murray95,
Proga00}. 
The lower limits of $ \dot{E_{\rm k}}/\ledd$ are 0.02\%-0.07\% which 
are substantially smaller than that is required to have significant 
feedback effect on their host galaxies.

3/58 of quasars with flat \civ~absorption troughs show spiky \lya~emission line, 
it may indicate that the small scale BALRs are rare.
Although the outflow radii of a number of individual 
quasars measured using density-sensitive absorption lines from excited levels
often find tens to 100-1000 pc scales, this contraversy has been long-standing
in this field \citep{Lucy14}, and no available data that facilitate a direct 
comparison for our quasar sample.

\section{ACKNOWLEDGMENTS}
Authors thank the anonymous referee for constructive
comments and suggestions for improving the clarity of the
manuscript.
We acknowledge the financial support by the Strategic Priority Research Program "The Emergence of Cosmological 
Structures" of the Chinese Academy of Sciences (XDB09000000), NSFC (NSFC-11233002, NSFC-11421303, U1431229) and 
National Basic Research Program of China (grant No. 2015CB857005).

Guilin Liu is supported by the National Thousand Young Talents Program of China, and acknowledges the grant from the National Natural Science Foundation of China (No. 11673020 and No. 11421303) and the Ministry of Science and Technology of China (National Key Program for Science and Technology Research and Development, No. 2016YFA0400700).
\label{sec:Result}


\end{document}